# Defect Engineering for Stabilizing Magnetic and Topological Properties in Mn(Bi$_{1-x}$Sb$_x$)$_2$Te$_4$

Haonan Chen[1#], Jiayu Wang[1#], Huayao Li[2#], Xunkai Duan[3, 4#], Yuxiang Wang[1], Zixuan Xu[1], Yingchao Xia[1], Wenhao He[5], Zehao Jia[5, 6], Xiangyu Cao[5, 6], Yicheng Mou[1], Xiangyu Jiang[7, 8, 9], Jiaming Gu[1], Pengliang Leng[1], Fengfeng Zhu[10, 11], Changlin Zheng[5], Xiang Yuan[7, 8, 9, 12], Faxian Xiu[1, 5, 6, 13, 14], Tong Zhou[3*], Lin Miao[2*], Cheng Zhang[1, 13*]

[1] State Key Laboratory of Surface Physics and Institute for Nanoelectronic Devices and Quantum Computing, Fudan University, Shanghai 200433, China
[2] Key Laboratory of Quantum Materials and Devices of Ministry of Education, School of Physics, Southeast University, Nanjing 211189, China
[3] Ningbo Institute of Digital Twin, Eastern Institute of Technology, Ningbo, Zhejiang 315200, China
[4] School of Physics and Astronomy, Shanghai Jiao Tong University, Shanghai 200240, China
[5] State Key Laboratory of Surface Physics and Department of Physics, Fudan University, Shanghai 200433, China
[6] Shanghai Qi Zhi Institute, 41st Floor, AI Tower, No. 701 Yunjin Road, Xuhui District, Shanghai 200232, China
[7] State Key Laboratory of Precision Spectroscopy, East China Normal University, Shanghai 200241, China
[8] Key Laboratory of Polar Materials and Devices, Ministry of Education, School of Physics and Electronic Science, East China Normal University, Shanghai 200241, China
[9] Shanghai Center of Brain-Inspired Intelligent Materials and Devices, East China Normal University, Shanghai 200241, China
[10] 2020 X-Lab, Shanghai Institute of Microsystem and Information Technology, Chinese Academy of Sciences, Shanghai 200050, China
[11] National Key Laboratory of Materials for Integrated Circuits, Shanghai Institute of Microsystem and Information Technology, Chinese Academy of Sciences, Shanghai 200050, China
[12] Chongqing Institute of East China Normal University, Chongqing 401120, China
[13] Zhangjiang Fudan International Innovation Center, Fudan University, Shanghai 201210, China
[14] Shanghai Research Center for Quantum Sciences, Shanghai 201315, China

[#] These authors contributed equally to this work
[*] Correspondence and requests for materials should be addressed to C. Z. (E-mail: zhangcheng@fudan.edu.cn), L. M. (E-mail: lmiao@seu.edu.cn), and T. Z. (E-mail: tzhou@eitech.edu.cn)

**Abstract**

MnBi$_2$Te$_4$ is a versatile platform for exploring diverse topological quantum states, yet its potential is hampered by intrinsic antisite defects. While Sb substitution has been employed to tune the Fermi level towards the charge neutral point, it exacerbates the formation of Mn-Sb antisite defects. Here, we address this challenge by combining first-principles calculations with strategic synthesis to systematically investigate and control antisite defects in Mn(Bi$_{1-x}$Sb$_x$)$_2$Te$_4$. Our calculations reveal that increasing antisite defect density progressively destroys the field-forced magnetic Weyl state, eventually driving the system into a trivial magnetic insulator. Motivated by these findings, we develop an optimized chemical vapor transport method, yielding high-quality Mn(Bi$_{1-x}$Sb$_x$)$_2$Te$_4$ crystals with significantly reduced antisite defect density. The emergence of strong Shubnikov-de Haas oscillations in the forced ferromagnetic state and a pronounced anomalous Hall effect near charge neutrality, with opposite signs for *n*- and *p*-type samples, confirms the type-II Weyl semimetal nature. These findings underscore the critical role of antisite defects in determining the magnetic and topological properties of Mn(Bi$_{1-x}$Sb$_x$)$_2$Te$_4$ and establish defect engineering via optimized synthesis as a crucial strategy for realizing its exotic magnetic topological states.

**Introduction**

MnBi$_2$Te$_4$ represents a unique material platform that facilitates comprehensive investigation of diverse topological quantum states and emergent quantum phenomena[1–3], spanning from antiferromagnetic (AFM) topological insulators[4–6] to ferromagnetic (FM) Weyl semimetals[7–10], and even theoretically predicted Möbius insulators within the canted antiferromagnetic (cAFM) phase[11]. In reduced dimensions, exfoliated MnBi$_2$Te$_4$ layers can host quantum anomalous Hall states[12,13], axion insulator states[14,15], as well as cAFM Chern insulator states[16], contingent on layer thickness and magnetic configuration[17]. Its inherent tunability, coupled with readily accessible magnetic transitions, positions it as an ideal system for probing the interplay between magnetism and topology. Indeed, the relatively low critical field of MnBi$_2$Te$_4$ enables facile transitions into field-forced FM Weyl semimetal states with a minimal set of Weyl points, approaching the long-sought paradigm of ideal Weyl semimetals[4,6].

Despite this promise, experimental progress has been limited by the presence of antisite defects, particularly Mn atoms occupying Bi sites and the reverse scenario[18–22]. These defects introduce additional intralayer AFM coupling, suppressing the saturation magnetization and disrupting the ideal Mn moment distribution within the lattice[22–26]. In the AFM topological insulator ground state, the misplaced Mn moments profoundly affect surface magnetic coupling, leading to a substantial and often debated reduction in the Dirac point gap size[20,21,27,28]. This has led to conflicting angle-resolved photoemission spectroscopy (ARPES) observations with both gapped and gapless surface states with significant sample-dependent variations[5,21,27,29–33]. Furthermore, antisite defects contribute to electron over-doping and reduced carrier mobility, shifting the Fermi level away from the desired topological regime and complicating the detection of Weyl points and Fermi arc states in transport measurements.

To mitigate these issues, Sb substitution has been widely adopted to form Mn(Bi$_{1-x}$Sb$_x$)$_2$Te$_4$, with the goal of lowering the Fermi level towards the energy window of the Weyl nodes[7,8,10,18,29,34]. However, this approach introduces a critical challenge: the chemical similarity between Sb and Mn unfortunately exacerbates the formation of Mn-Sb antisite defects[18]. Consequently, while Sb doping can optimize carrier density, the resulting increase in antisite concentration may simultaneously degrade both magnetic and topological properties. This trade-off obscures the intrinsic characteristics of Mn(Bi$_{1-x}$Sb$_x$)$_2$Te$_4$, hindering unambiguous interpretation of its fundamental properties. A central question thus remains: can the benefits of Fermi level tuning be retained while suppressing the detrimental effects of defect-induced disorder? Addressing this requires a clear understanding of how antisite defects influence magnetic and topological behavior, along with effective methods to control their formation during synthesis.

In this work, we systematically investigate the role of antisite defects in Mn(Bi$_{1-x}$Sb$_x$)$_2$Te$_4$ through first-principles calculations and targeted materials synthesis. Our calculations reveal that increasing defect concentrations progressively degrade the Weyl cone, ultimately leading to a trivial magnetic insulating phase. Given the intimate interplay between band topology and magnetism in this system, we attribute the topological degradation to the antisite-induced suppression of the field-forced FM state (Weyl semimetal), also the AFM ground state (topological insulator). This finding emphatically underscores the imperative to minimize antisite defects to access the intrinsic topological properties of this material. Motivated by these findings, we develop an optimized chemical vapor transport (OCVT) growth method that enables precise control of growth conditions, including temperature profiles adapted to Sb content and an excess of Mn and Te precursors. This approach yields high-quality, stoichiometric Mn(Bi$_{1-x}$Sb$_x$)$_2$Te$_4$ crystals with markedly improved quality, representing a significant advancement over conventional self-flux and solid-state reaction approaches.

Experimental characterizations of OCVT-grown samples confirm the success of this strategy. Magnetization measurements show enhanced AFM order in the ground state, characterized by elevated Néel temperature and enhanced saturation magnetization, indicative of a significant reduction in antisite defect density. ARPES and transport studies reveal a systematic tuning of the Fermi level from $n$- to $p$-type with increasing Sb content. Notably, at an optimized doping level of $x = 0.20$, we achieve a minimum carrier density ($7 \times 10^{17}$ cm$^{-3}$) and a record-high carrier mobility (2519 cm$^2$/V·s). In the field-forced FM state, strong Shubnikov-de Haas oscillations appear, with oscillation frequencies tracking carrier density. A pronounced anomalous Hall effect (AHE) emerges as the Fermi level nears charge neutrality, with opposite anomalous Hall conductivity signs for $n$- and $p$-type samples, further corroborating the type-II Weyl semimetal character. Together, these results emphasize the critical and previously underappreciated role of antisite defects in determining the magnetic and topological properties of Mn(Bi$_{1-x}$Sb$_x$)$_2$Te$_4$. By combining theoretical insights with advanced synthesis, this work establishes defect engineering as an essential strategy for realizing and stabilizing exotic magnetic topological phases in this material family.

## Results

**Antisite defects driven degradation of magnetism and band topology.** MnBi$_2$Te$_4$ crystallizes in the space group $R$-3m (No.166)[35], consisting of stacked septuple layers (SLs) that are weakly bonded through van der Waals interaction along the $c$ axis. Within each SL, atoms are arranged in a Te-Bi-Te-Mn-Te-Bi-Te sequence (Fig. 1a), forming a triangular lattice in the $ab$ plane with $C_3$ rotation symmetry.

During crystal growth, MnBi$_2$Te$_4$ inevitably forms point defects, including vacancies (denoted as V$_A$), interstitials (A$_i$), and crucially, antisites (A$_B$, denoting A occupying B site), where A and B represent different atomic species[18,20]. These defects act as electron acceptors or donors, significantly modulating the Fermi level. While ideal MnBi$_2$Te$_4$ is predicted to be a topological insulator with a Fermi level within the bandgap, as-grown crystals consistently exhibit pronounced *n*-type doping, as confirmed by Hall effect and ARPES[5,8,14,24,27,29]. First-principles calculations suggest that the antisites Bi$_{Mn}$ and Mn$_{Bi}$, arising from lattice mismatch between Bi$_2$Te$_3$ and MnTe building blocks, possess the lowest formation energies, making them the most prevalent defects[18,20]. The predominant presence of Mn$_{Bi}$ and Bi$_{Mn}$ antisites in MnBi$_2$Te$_4$ has been further verified experimentally[24,36–39]. Vacancies, interstitials and Bi$_{Te}$ antisites, having higher formation energies, are comparatively less abundant. Schematics of these defect types are illustrated in Fig. 1a. Notably, Mn$_{Bi}$ antisites act as electron-acceptors, introducing holes, while the Bi$_{Mn}$ antisites are electron donors[18]. Quantitative analysis (see Supplementary Table 1) indicates that non-stoichiometric crystals typically contain more Bi$_{Mn}$ than Mn$_{Bi}$, resulting in the observed heavy *n*-doped conduction in MnBi$_2$Te$_4$, with the electron density of approximately $10^{20}$ cm$^{-3}$[8,14,24,29].

Furthermore, these prevalent antisite defects (Bi$_{Mn}$ and Mn$_{Bi}$) exert a substantial influence on the magnetic configuration of MnBi$_2$Te$_4$ across varying magnetic fields. Below the Néel temperature ($T_N$), Mn spins are expected to adopt an A-type AFM ground state[5]. In the idealized scenario, Mn atoms are localized exclusively within the central Mn layer of each SL, denoted as Mn$_{Mn}$, with their moments exhibiting intralayer ferromagnetic order. These ferromagnetic SLs couple antiferromagnetically with one another along the *c* axis (denoted by the red arrows in Fig. 1b). However, excessive Mn$_{Bi}$ antisites result in a redistribution of spin moments into the nominally non-magnetic Bi layers (the orange arrows in Fig. 1b), where Mn$_{Bi}$ antisites develop an antiparallel ordering relative to the Mn$_{Mn}$ moments. This induces a partially compensated net magnetization and establishes a ferrimagnetic (FiM) order for the whole SL, attributable to the strong AFM coupling between the non-equivalent Mn moments[24,26]. Meanwhile, Bi$_{Mn}$ antisites introduce magnetic dilution within the Mn layers. Upon application of an external magnetic field exceeding the saturation field, the Mn$_{Mn}$ moments are forced to align with the field, establishing a FM order. The Mn$_{Bi}$ antisites, however, continue to compensate the magnetic moments due to the strong AFM intralayer coupling[26]. Consequently, the saturation magnetization $M_{sat}$ at moderate fields (~ 9 T) is suppressed owing to both compensation by Mn$_{Bi}$ antisites and dilution from Bi$_{Mn}$ antisites (Fig. 1c), causing the value inconsistencies in experimentally obtained $M_{sat}$[8,26,29,34]. Overcoming intralayer AFM coupling, thereby achieving full alignment of all Mn moments across all layers, necessitates much higher magnetic fields, experimentally demonstrated to be around 50 T (Fig. 1d)[26].

Beyond their impact on magnetic ordering, Mn$_{Bi}$ and Bi$_{Mn}$ antisites fundamentally alter the band topology. While prior theoretical calculations primarily focused on idealized ferromagnetic order, assuming perfect Mn moment alignment in defect-free crystals, these investigations predicted a single pair of type-II Weyl points in field-forced FM MnBi$_2$Te$_4$[4,6,7,10,40,41]. They also suggested potential topological phase transitions to type-I Weyl semimetal or trivial magnetic insulator states with minor strain or lattice parameter variations (~ 1%)[6]. The susceptibility of the Weyl structure in MnBi$_2$Te$_4$ indicates the necessity of investigating the defect-induced topological transitions. Standard magnetotransport measurements, typically conducted below 12 T, operate in a regime where antisite defects induce a FiM order and introduce additional magnetic sites (Fig. 1c), rendering the idealized FM state assumption questionable. We calculated the band structure as a function of antisite density (Figs. 1e-g), considering the coexistence of Mn$_{Bi}$ and Bi$_{Mn}$ antisites.

At an antisite density of 6.25% $Mn_{Bi}$ and 12.5% $Bi_{Mn}$ (Fig. 1f), band crossing is lifted, opening a gap that further enlarges with the increase of defect concentration (Fig. 1g). Calculation details and defective crystal structures are summarized in Supplementary Figs. 1 and 2. A schematic of defect-induced band topology evolution is shown in Fig. 1h, demonstrating a topological phase degeneration from ideal Weyl semimetals to trivial insulators.

The strategic substitution of Bi with Sb atoms provides an effective mechanism for Fermi level engineering towards charge neutrality[29,34], a critical prerequisite for accessing the emergent phenomena dominated by Weyl fermions[7–10]. Specifically, Sb-doping augments both $Mn_{Bi/Sb}$ and $Bi/Sb_{Mn}$ defect concentrations, while $Mn_{Bi/Sb}$ defects show a more pronounced increase[18], leading to a complex interplay but generally contributing to a reduction in Fermi energy. However, inadvertently introduced deleterious $Mn_{Bi/Sb}$ and $Bi/Sb_{Mn}$ antisites further disrupt the inherent magnetism[22,23,25] by inducing chaotic Mn moment distribution. Energy calculations across various magnetic configurations and defect densities (Supplementary Note 3) reveal that both antisite types cooperatively suppress interlayer AFM coupling and promote undesired intralayer antiparallel alignment. Further calculations emphasize the significance of complete FM alignment in band crossing and Weyl cone formation, which is strongly hindered by the strengthened AFM intralayer coupling at higher antisite densities (Supplementary Fig. 4). These excessive antisites, altering the magnetic order to a large extent, profoundly degrade the band topology in the FM $Mn(Bi_{1-x}Sb_x)_2Te_4$ family (Figs. 1e-h), hindering access to Weyl-point-driven phenomena. Therefore, it's necessary to tune the Fermi level towards the Weyl points, and to concurrently minimize antisite defect density for recovering the intrinsic magnetism and band topology.

**OCVT synthesis.** As these defects are inherent to the crystal growth process and their concentration is intimately linked to growth dynamics, optimizing synthesis methods is critical for defect reduction, thereby unlocking the intrinsic magnetic and topological properties of $Mn(Bi_{1-x}Sb_x)_2Te_4$. While the conventional chemical vapor transport approach has shown promise for improving chemical stoichiometry in $MnBi_2Te_4$ growth[39,42], its application to $Mn(Bi_{1-x}Sb_x)_2Te_4$ may paradoxically result in higher defect densities compared to self-flux grown crystals[42]. Here, we introduce an optimized chemical vapor transport (OCVT) methodology (Fig. 2a) meticulously engineered to synthesize stoichiometric $Mn(Bi_{1-x}Sb_x)_2Te_4$ with substantially suppressed antisite defects, directly addressing the limitations of conventional methods. This approach achieves precise control of Fermi level and carrier density through systematic compositional tuning of Sb content; in the meantime, the remarkable reduction in antisite density effectively stabilizes both the intrinsic magnetic structure and nontrivial band topology in $Mn(Bi_{1-x}Sb_x)_2Te_4$.

In detail, recognizing the inherent non-stoichiometry of Mn observed in self-flux samples and the slow transportation efficiency of Mn, our OCVT strategy hinges on maintaining a precise excess of Mn in the starting materials, coupled with rigorous pre-heating protocols to ensure stoichiometric Mn incorporation. Critically, introducing additional Mn necessitates a corresponding increase in Te to create a Te-rich growth environment. This Te-rich condition is instrumental in minimizing the formation of $Bi/Sb_{Mn}$ and $Bi/Sb_{Te}$ defects[18], while the merely controlled Mn excess preferentially leads to the benign precipitation of MnTe. Based on this strategy, we mixed Mn, Bi, Sb and Te elements in a ratio of MnTe: $(Bi, Sb)_2Te_3$ = $y$: 1 ($y$ ranging from 2 to 3). The raw material mixture was then flame-sealed in vacuum quartz ampules with $I_2$ as the transport agent and placed in a two-zone tube furnace. Considering the elevated melting point and crystallization temperature of Mn(Bi$_{1-x}$Sb$_x$)$_2$Te$_4$ with higher Sb content, it is crucial to use higher temperatures at both the hot and cold ends of the quartz ampule (Fig. 2b). With a small temperature gradient of approximately 1 °C/cm, I$_2$ agent transport the material towards the cold ends. Detailed growth conditions are listed in Supplementary Table 6. After a 14-day reaction, we obtained shiny hexagonal or rectangular plate-shaped crystals at the cold ends (the optical images are shown in Supplementary Figure 5).

To assess the structural impact of Sb doping, X-ray diffraction (XRD) measurements were performed on the (001)-oriented crystal across a range of Sb concentrations ($x$ = 0 to 0.26), as shown in Supplementary Fig. 6a. Analysis of the XRD peak positions reveals a largely invariant $c$-lattice parameter across the doping range (see Supplementary Fig. 6b), consistent with prior reports on lightly Sb-doped MnBi$_2$Te$_4$[29,34] and indicating that modest Sb substitution effectively preserves the parent crystal structure. Energy-dispersive X-ray (EDX) spectroscopy confirmed the near-stoichiometric elemental composition of these OCVT-grown crystals (see Table 1 and Supplementary Figs. 7–8), demonstrating the efficacy of our optimized synthesis. All the $x$ values in this work are the actual Sb content, determined through EDX by $x = \text{Sb}_{\text{EDX}}\%/(\text{Bi}_{\text{EDX}}\% + \text{Sb}_{\text{EDX}}\%)$. Scanning transmission electron microscopy (STEM) and scanning tunneling microscopy (STM) measurements were also performed on the OCVT-grown crystals (see Supplementary Figs. 10–11), confirming a high-quality crystallinity.

**Enhanced magnetism properties due to reduced defect density.** To rigorously assess the impact of our OCVT synthesis on mitigating defect-induced magnetic degradation (shown in Figs. 1b-d), we performed comprehensive magnetization measurements to both qualitatively and quantitatively evaluate defect concentrations in OCVT-grown Mn(Bi$_{1-x}$Sb$_x$)$_2$Te$_4$ crystals. Figure 3a presents temperature-dependent magnetic susceptibility curves, $\chi(T)$, of Mn(Bi$_{1-x}$Sb$_x$)$_2$Te$_4$ across the doping range $x$ = 0 to 0.26, acquired under field-cooled (FC) conditions with **H** // $c$. The near-perfect overlap between zero-field-cooled (ZFC) and FC curves (see Supplementary Fig. 12a) signifies the homogeneity and high quality of our OCVT-grown crystals. All $\chi(T)$ curves exhibit consistent AFM behavior, characterized by a distinct peak corresponding to $T_N$. Below $T_N$, the magnetization sharply decreases towards zero upon further cooling, unequivocally evidencing robust long-range AFM ordering. The absence of any signature indicative of FiM behavior in all $\chi(T)$ curves underscores the successful suppression of magnetic defects or phase impurity such as Mn-doped (Bi, Sb)$_2$Te$_3$[24,43]. As anticipated, $T_N$ systematically decreases with increasing Sb content $x$, reflecting the weakening of interlayer AFM coupling due to the increasing, albeit minimized, density of Mn$_{\text{Bi/Sb}}$ and Bi/Sb$_{\text{Mn}}$ antisites even in optimized growth.

Figure 3c illustrates representative magnetization versus magnetic field $M(H)$ for $x$ = 0.04. As the magnetic field increases, the AFM order transitions to cAFM state via spin-flopping at $\mu_0 H_{c1}$ ~ 3.8 T. Upon further field increase, the cAFM phase transitions to a FM state at the saturation field $\mu_0 H_{c2}$ ~ 8.2 T. The magnetization saturates to $M_{\text{sat}}$ within 9 T ($M_{\text{sat}}$ refers to the magnetization value in the configuration of Fig. 1c). Critical fields $H_{c1}$ and $H_{c2}$, identified by peak and valley positions in the first and second derivatives of magnetization (d$M$/d$H$ and d$^2M$/d$H^2$, detailed in Supplementary Fig. 13) respectively, are shown to decrease at elevated temperatures, with $M(H)$ behavior approaches the paramagnetic limit (Fig. 3c). A comprehensive magnetic phase diagram summarizing these transitions is presented in Fig. 3e. $M(H)$ curves at 2 K for each composition are presented in Fig. 3b, revealing a consistent magnetization trend (AFM→cAFM→FM). Both spin-flop and saturation critical fields exhibit a systematic decrease with increasing Sb content (Fig. 3d). Detailed analysis of $M(H)$ across different temperatures and Sb concentrations allows us to construct a three-dimensional magnetic phase diagram (Fig. 3f), mapping distinct magnetic phases across the entire doping range and temperature domains. Additional magnetization curves at various temperatures and torque magnetometry for samples from OCVT are shown in Supplementary Figs. 14 and 15, respectively.

Figure 4a compares $T_N$ as a function of Sb content $x$ from this study and previously reported values. $T_N$ values, extracted from kinks of $\chi(T)$ and longitudinal resistivity $\rho_{xx}(T)$ curves (see Supplementary Fig. 13), reflect the strength of interlayer AFM coupling among SLs against thermal fluctuations as a fundamental indicator. Higher $T_N$ suggests a more robust AFM ground state with fewer antisite defects. However, with increasing Sb content and consequently, a higher density of $Mn_{Bi/Sb}$ and $Bi/Sb_{Mn}$ antisites (even in optimized growth), the interlayer AFM interaction weakens, lowering the energy barrier towards FiM ground states. In the Sb-rich limit of $MnSb_2Te_4$, both FiM and AFM ground states have been observed depending on the $Mn_{Sb}$ concentration resulting from different growth conditions, with Curie temperatures ($T_C$) ranging from 24 K to 58 K or a lower $T_N$ of around 19 K[19,22,23,34,44–46].

As shown in Fig. 4b, $M_{sat}$ decreases nearly linearly with a rising Sb content due to the compensation between nonequivalent Mn moments caused by increasing $Mn_{Bi/Sb}$ antisites, while the $Bi/Sb_{Mn}$ antisites act as effective magnetic vacancies within Mn layers, diluting magnetic ion concentration and thereby reducing overall magnetization. In $Mn(Bi_{1-x}Sb_x)_2Te_4$, Mn exists in a high-spin $Mn^{2+}$ state ($S = 5/2$), yielding an expected saturation magnetization of 5.0 $\mu_B/Mn^{2+}$ according to the Hund's rule ($\mu_B$ being the Bohr magneton). Previous neutron diffraction, high-field magnetization and theoretical calculations for $MnBi_2Te_4$ consistently report a local magnetic moment of 4.04 ~ 4.7 $\mu_B/Mn^{2+}$[5,24–26,34,47], with minor deviations potentially arising from orbital hybridization[26]. Importantly, within a 9 T field, the sublattices of $Mn_{Mn}$ and $Mn_{Bi/Sb}$ form FM configuration respectively yet remain antiparallel to each other (see Fig. 1c). Consequently, the $M_{sat}$ observed at 9 T is reduced by the $Mn_{Bi/Sb}$ moments and can be effectively used to estimate the concentration of Mn-Bi/Sb mixing[26,39].

To quantify defect concentrations, we set the amount percentage of $Mn_{Mn}$ and $Mn_{Bi/Sb}$ in corresponding atom layers as $a$ and $b$, respectively. Considering the different FM configurations shown in Figs. 1c and 1d, we write:

$$a + 2b = Mn_{EDX} \quad (1)$$

$$\frac{a - 2b}{a + 2b} = \frac{M_{sat}}{M_0} \quad (2)$$

where $Mn_{EDX}$ is the normalized Mn proportion obtained from experiments with $Te_{EDX}$ as a reference value of 4. $M_0$ denotes the local magnetic moment of $Mn^{2+}$, which can be approximated to be 4.6 $\mu_B/Mn^{2+}$, the fully saturated magnetization in experiments at 50 T[26]. Neglecting high-energy vacancies $V_{Mn}$, the concentration of $Bi/Sb_{Mn}$ can be approximated as $(1 - a) \times 100\%$. Calculations based on Equations (1) and (2) yield Sb-dependent defect concentrations summarized in Table 1. Compared to $Mn(Bi_{1-x}Sb_x)_2Te_4$ synthesized via other methods, our OCVT-grown crystals show fewer $Mn_{Bi/Sb}$ and $Bi/Sb_{Mn}$ defects, directly correlating with the observed enhancement in magnetic properties and the potential for realizing intrinsic topological states. The clear negative correlation between $b$ and $M_{sat}$ further underscores the low defect density achieved in our samples, corroborated by the $M_{sat}$ data from other works shown in Fig. 4b.

**Tuning the Fermi level by Sb doping.** Beyond magnetic property refinement, substituting Bi with Sb in MnBi$_2$Te$_4$ provides a potent means to precisely tune the Fermi level, promoting a transition from *n*- to *p*-type conduction. Controlled doping in topological insulators is a crucial strategy for positioning the Fermi level within the bulk band gap while preserving topologically protected surface states, as exemplified by Sn-Bi$_{1.1}$Sb$_{0.9}$Te$_2$S[48]. This Fermi level tunability arises from the dopant role in compensating intrinsic defects by introducing counter-doping carriers.

To experimentally visualize this carrier-type transition, we performed ARPES measurements on OCVT-grown Mn(Bi$_{1-x}$Sb$_x$)$_2$Te$_4$ with *x* ranging from 0.14 to 0.26. Figures 5a-b present the ARPES spectra and their second derivative images, respectively. Sample cleavage and signal detection were both performed at 10 K below $T_N$. Higher than typically used for detecting topological surface states, the utilized photon energy of 13 eV reveals primarily the dispersion of bulk bands, to access optimal signal clarity[29]. Subtle indications of Dirac-like surface states appear only in the second derivative spectra. This limited surface signal is likely due to a Sb-induced gap opening at the Dirac point[49,50], as well as the surface inhomogeneity induced scattering[51]. Unlike the minimal surface gap in MnBi$_2$Te$_4$, a recent work showed that the surface gap continuously grows with increasing Sb content, reaching 104 meV at 10 K for *x* = 0.1, comparable to the non-trivial bulk gap[49]. It may cause bulk and fully gapped surface states to overlap in the spectra. In our result, a clear downward shift of Fermi level $E_F$ occurs as *x* rises. At *x* = 0.14, $E_F$ is well above the bulk band gap (*E* - $E_F$ ~ -160 meV), indicating *n*-type carriers. At *x* = 0.20, $E_F$ approaches the conduction band minimum (CBM), while at *x* = 0.22, it enters the valence band and locates near the valence band maximum (VBM), signaling a shift to *p*-type conduction. At *x* = 0.26, a second valence band becomes visible. Notably, to confirm the band edge proximity of the Fermi level, two samples with *x* = 0.20 were measured, exhibiting few differences in the $E_F$. Indeed, the almost insulating bulk state at *x* ~ 0.20 makes the $E_F$ more sensitive to even minor carrier variations.

Collectively, ARPES data demonstrate a monotonic downward shifting of $E_F$ from the conduction band to the valence band with increasing Sb doping, indicating the *n*-to-*p* carrier transition, with a charge neutral point (CNP) at *x* ~ 0.20.

**Transport properties of Mn(Bi$_{1-x}$Sb$_x$)$_2$Te$_4$.** To further investigate the *n-p* carrier evolution, we performed electrical transport measurements of OCVT-grown Mn(Bi$_{1-x}$Sb$_x$)$_2$Te$_4$ crystals. Temperature-dependent longitudinal resistivity $\rho_{xx}(T)$ curves, measured from 300 K down to 2 K, exhibit semimetallic behavior across all doping (Fig. 6a). Notably, samples with *x* = 0.19 to 0.21, corresponding to Fermi level proximity to the band edge, show elevated resistivity values across the entire temperature range, consistent with reduced carrier density near the charge neutrality point. The *x* = 0.20 sample, closest to the CNP, displays a nonmonotonic $\rho_{xx}(T)$ behavior with a distinct upturn around 165 K. A kink in $\rho_{xx}(T)$ near 24.5 ~ 26 K marks the Néel temperature, indicative of a sharp change in spin scattering as magnetic moments order below $T_N$[34].

Field-dependent magnetoresistance (MR) and Hall resistivity ($\rho_{yx}$) measurements at 2 K (Figs. 6b-c) reveal further transport characteristics (magnetic field applied along the *c* axis). MR, defined as [$\rho_{xx}(H)$-$\rho_{xx}(0)$]/$\rho_{xx}(0)$×100%, exhibits distinct slope changes at critical fields $H_{c1}$ and $H_{c2}$, corresponding to magnetic phase transitions. Here, $\rho_{xx}(0)$ refers to the zero-field resistivity. Hall resistivity variations across magnetic phases demonstrate an AHE contribution, suggesting the coupling between band structure and magnetism. Beyond $H_{c2}$, where magnetization saturates, the anomalous Hall term stabilizes. To quantitatively determine carrier density, we analyzed Hall data exclusively in FM states by extracting the linear slope of $\rho_{yx}$ versus magnetic field, defined as $R_H = d\rho_{yx}/d\mu_0 H$, where $n = 1/(e \cdot R_H)$ (negative values for electrons, positive for holes). As $x$ increases from 0 to 0.23, $R_H$ shifts from negative to positive, directly confirming the $n$-to-$p$ carrier type transition. Pronounced nonlinearity in the $\rho_{yx}$ curve for $x = 0.20$, contrasting sharply with the near-linear behavior for slightly $n$-doped ($x = 0.19$) and $p$-doped ($x = 0.21$) samples, suggests two-band conduction with significant contributions from both electrons and holes near CNP. To extract reliable carrier densities, we employed a two-band model to fit the nonlinear Hall data. Due to the limit of ±12 T field range in our experiments, fitting FM state Hall signals may be error-prone, as only a limited range (~ 4 T) is suitable for fitting. To minimize errors, we used nonlinear Hall data above $T_N$ for more reliable fits, where the magnetization-related anomalous Hall contributions are negligible in the paramagnetic state. This approach yields a carrier density down to around $7 \times 10^{17}$ cm$^{-3}$ through fitting (details in Supplementary Fig. 16).

Carrier mobility $\mu$, another important transport metric, calculated as $\mu = R_H/\rho_{xx}(0)$, exhibits a clear correlation with carrier density (Fig. 6d). Samples with lower carrier densities exhibit higher mobilities, with mobility peak and density valley coinciding around $x = 0.20$, further corroborating the $n$-$p$ transition near the CNP observed in ARPES. Remarkably, the maximum mobility reaches 2519 cm$^2$/V·s, significantly exceeding previously reported values for both pristine MnBi$_2$Te$_4$ and Mn(Bi$_{1-x}$Sb$_x$)$_2$Te$_4$ (comparison shown in Table 2), underscoring the efficacy of our OCVT synthesis for high-quality Mn(Bi$_{1-x}$Sb$_x$)$_2$Te$_4$. Carrier densities and mobilities of all measured samples via OCVT are summarized in Supplementary Fig. 17, indicating an overall enhancement of carrier mobilities.

Crucially, the Sb doping level corresponding to the charge neutrality point is not a fixed material parameter but rather synthesis-dependent, varying significantly across growth methods. Prior studies report CNP compositions mostly ranging from $x = 0.25$ to $0.35$[7–10,29,34,43,52,53], whereas OCVT-grown crystals exhibit a pronounced leftward shift of the CNP to a lower Sb content, around $x = 0.20$ (see Supplementary Fig. 18). This CNP shift is a compelling consequence of the reduced defect density achieved via OCVT synthesis. As evidenced by magnetization data in Table 1, the lower Mn$_{Bi/Sb}$ concentration in OCVT-grown crystals effectively stabilizes the intrinsic AFM order and, simultaneously, shifts the Fermi level upward. Compared to samples grown via other methods, OCVT-grown ones exhibit fewer electron-donating Bi/Sb$_{Mn}$, whose quantity reduces even more than Mn$_{Bi/Sb}$ (see Table 1 and Supplementary Table 1), thereby generally enhancing $p$-doping efficiency for a given Sb concentration and resulting in a CNP at lower Sb doping levels. This capability to reach the CNP at a reduced Sb concentration directly translates to a significant suppression of antisite defects as the Fermi level approaches towards Weyl points. As directly visualized in Fig. 6e, at the optimal doping contents close to the CNP, samples from OCVT exhibit superior stability of the intrinsic magnetic order against antisite-induced disturbance, characterized by concurrent maxima in both $M_{sat}$ and $T_N$. The optimized magnetism profoundly suppresses the defect-induced degradation on the intrinsic band topology, thereby preserving the pristine topological characteristics while facilitating reliable access to key emergent phenomena governed by Weyl points or nontrivial surface states inherent to stoichiometric Mn(Bi$_{1-x}$Sb$_x$)$_2$Te$_4$.

**Quantum oscillations and anomalous Hall effect in Sb-doping driven electronic structure**

**evolution.** Pronounced Shubnikov-de Haas (SdH) oscillations emerge in both $\rho_{xx}$ and $\rho_{yx}$ when Mn(Bi$_{1-x}$Sb$_x$)$_2$Te$_4$ bulks are driven into forced FM states by a magnetic field along the $c$ axis. After subtracting the background with polynomial fitting, the oscillation term $\Delta\rho$ as a function of $1/\mu_0 H$ is shown in Fig. 7a. For visual clarity, oscillations are separated by 0.2 mΩ·cm increments across samples with the carrier density ranging from $n$- to $p$-type. The frequency of oscillations $F_S$, derived through fast Fourier transform (FFT) analysis, are shown in Fig. 7b. The $F_S$ values, linearly related to the Fermi pocket size $A_F$ perpendicular to the field direction via Onsager's relation, $F_S = \hbar/2\pi e\, A_F$, reflect the Fermi level position and align with the observed carrier density values. Figure 7c shows the relationship between $F_S$ and carrier density $n$, where $F_S$ decreases as carrier density declines on both electron and hole-doped sides. Notably, these SdH oscillations originate from bulk FM-state bands rather than topological surface states. Previous high-field SdH oscillations up to 60 T confirmed that the ferromagnetic Mn(Bi$_{1-x}$Sb$_x$)$_2$Te$_4$ realizes an ideal type-II Weyl semimetal, with a pronounced difference in the three-dimensional Fermi pocket shape between highly $n$-doped and $p$-doped samples[7,10]. To analyze the Fermi surface shape and anisotropy in $n$- and $p$-doped Mn(Bi$_{1-x}$Sb$_x$)$_2$Te$_4$, we approximate the Fermi pocket as an ellipsoid with axes $k_a = k_b = \gamma k_c$ and apply Onsager's relation to estimate the carrier density across three-dimensional ellipsoidal Fermi surfaces using the following formula:

$$F_S = \frac{\hbar}{2\pi e} A_F = \frac{\hbar}{2e}\gamma^2 k_c^2 \tag{3}$$

$$n_{3D} = \frac{2}{(2\pi)^3}\frac{4\pi}{3}k_a k_b k_c = \frac{1}{3\pi^2}\gamma^2 k_c^3 \tag{4}$$

Thus, the relationship between $F_S$ and carrier density $n$ can be expressed as:

$$F_S = \frac{\hbar}{2e}(3\pi^2 n\gamma)^{\frac{2}{3}} \tag{5}$$

The factor $\gamma$ can be extracted from the fitting procedure, which reflects the Fermi surface anisotropy. We can get $k_a = k_b = 0.37\, k_c$ for the electron-doped regions and $k_a = k_b = 0.107\, k_c$ for the hole-doped regions. This pronounced difference in Fermi surface anisotropy between conduction and valence bands is fully consistent with the characteristic tilted Weyl cone dispersion of a type-II Weyl semimetal.

Temperature-dependent SdH oscillations (Fig. 7d) persist up to temperatures exceeding $T_N$, with their amplitude diminishing upon warming, attributable to spin fluctuations that allow the magnetic field to induce a moderate alignment of Mn$^{2+}$ moments even within the paramagnetic phase, consistent with the previous SdH oscillations and electron spin resonance (ESR) characterizations[5,8–10,54]. From the thermal damping of SdH oscillations, we estimate the effective mass value $m^* = 0.06\, m_0$ (where $m_0$ is the electron mass) by fitting the temperature dependence of FFT amplitude with Lifshitz-Kosevich (L-K) formula (Inset in Fig. 7d):

$$R_T = \frac{\alpha T m^*}{B\sinh(\alpha T m^*/B)} \tag{6}$$

where $R_T$ refers to the amplitude of oscillations at particular $1/B$ or the amplitude of FFT analysis, and $\alpha=2\pi^2 k_B m_0/e\hbar$. The magnetic field value in Equation (6) is taken as $1/B=(1/B_1+1/B_2)/2$, where $B_1$ and $B_2$ are the starting and ending fields of FFT analysis. Additionally, as temperature increases, shifts in SdH peak and valley positions reflect the strong coupling between the band structure and magnetic states (Fig. 7d and Supplementary Fig. 20).

The AHE, characterized by corresponding conductivity $\sigma_{AHE}$, provides further compelling transport evidence for the Weyl cone topology. The presence of a single Weyl point pair in ferromagnetic Mn(Bi$_{1-x}$Sb$_x$)$_2$Te$_4$ simplifies AHE analysis, avoiding complexities arising from multiple Weyl pairs or trivial bands. Theory predicts that type-I Weyl semimetals with broken time-reversal-symmetry exhibit Fermi level-independent $\sigma_{AHE}$ due to the complete compensation of Berry curvature from free carriers, only determined by the Weyl point separation $\Delta k$ in momentum space[55]. In contrast, type-II Weyl semimetals, characterized by tilted Weyl cones, generate a non-zero Berry curvature integral, yielding significant free-carrier contributions to AHE along with the $\Delta k$-related contributions[56]. The sign reversal of the Berry-curvature-induced term in $\sigma_{AHE}$, occurring as the Fermi level traverses Weyl points accompanied by shrinkage or emergence of electron/hole pockets, serves as the key criterion distinguishing type-II from type-I Weyl semimetals[7,8].

Figure 7e presents field-dependent anomalous Hall resistivity $\rho_{AHE}$ for samples with varied carrier densities, extracted via subtracting the ordinary Hall background from total $\rho_{yx}$. Across all samples, the $\rho_{AHE}$ values exhibit distinct evolution among different magnetic phases and gradually saturate in the FM state. Notably, $\rho_{AHE}$ changes its sign from negative to positive with Fermi level downshifting due to Sb doping. Furthermore, in both electron and hole doping regions, $|\rho_{AHE}|$ increases as Fermi level moving towards CNP. Converted into conductivity form, $\sigma_{AHE}$ values are strongly related to the carrier types and densities (see Fig. 7f). Figure 7g summarizes the anomalous Hall angle $\Theta_{AHE} = \sigma_{AHE}/\sigma_{xx}$ versus carrier density, further highlighting the Fermi level dependence. Similarly, the absolute value of $\Theta_{AHE}$ increases with decreasing carrier density for both electrons and holes within a moderate doping range, and abruptly reverses its sign as the Fermi level crosses the CNP. This characteristic evolution of $\Theta_{AHE}$ aligns closely with the expectations for an ideal type-II Weyl semimetal[7,8,56]. Moreover, the remarkably large $\Theta_{AHE}$ (~ 33.6%) observed near the CNP, along with the enhanced carrier mobility, underscores the dominant role of Weyl fermions in bulk transport. This enhanced anomalous Hall angle, arising from the intrinsic tilted Weyl cone and minimized contributions from irrelevant electronic states due to Fermi level tuning near charge neutrality, further solidifies the type-II Weyl semimetal nature in Mn(Bi$_{1-x}$Sb$_x$)$_2$Te$_4$ and the superior quality of OCVT-grown crystals in revealing these exotic quantum phenomena.

**Discussion**

Through strategic Sb-doping engineering, we systematically navigated the electronic landscape of Mn(Bi$_{1-x}$Sb$_x$)$_2$Te$_4$ as the Fermi level was progressively tuned across the CNP toward Weyl point proximity. However, this delicate Fermi level manipulation is inextricably linked to the unintended introduction of Mn$_{Bi/Sb}$ and Bi/Sb$_{Mn}$ antisite defects, which, as we have decisively demonstrated, exert a profound influence on the intrinsic magnetism and band topology. Moreover, Sb substitution itself introduces complexities, impacting the band structure through (i) continuous contraction of lattice parameters $a$ (= $b$) and $c$ with increasing Sb concentration[25,34], and (ii) reduced SOC strength of Sb atoms compared to Bi. The latter results in a diminished topological bulk bandgap within AFM ground states, which induces a topological transition to a trivial insulator at heavy doping levels ($x > 0.55$)[29]. These intertwined effects of Sb doping – Fermi level tuning and structural/electronic modulations – gain paramount importance considering the documented exquisite sensitivity of the type-II Weyl semimetal phase in FM MnBi$_2$Te$_4$ to both SOC strength and lattice parameters.

To dissect the isolated impact of Sb substitution itself on the Weyl cone topology, we further carried out first-principles calculations of the evolution of electronic structure for defect-free FM configurations upon Sb doping (calculation models in Supplementary Fig. 3). As shown in Figs. 8a-d, with the increase of Sb content, the Weyl points shift towards the Γ point with the separation Δ$k$ reducing. The comprehensive doping-dependent evolution of Weyl cone characteristics is summarized in Fig. 8e, a similar trend has been reported[57]. As mentioned above, higher Sb content naturally introduces more Mn$_{Bi/Sb}$ defects, further degrading the Weyl cone. However, the OCVT synthesis we developed effectively minimizes defect density for a given doping level, enabling Fermi level tuning near the CNP with minimal Sb doping ($x$ = 0.20). This methodological advancement enables reliable exploration of Weyl point-governed quantum transport phenomena.

These findings offer compelling new insights into the experimental realization and investigation of the ideal Weyl semimetal with minimum Weyl points in Mn(Bi$_{1-x}$Sb$_x$)$_2$Te$_4$. Such systems, achieved or predicted in only a select few magnetic materials like (Cr, Bi)$_2$Te$_3$[58], FM-Eu$_3$In$_2$As$_4$[59–62] and K$_2$Mn$_3$(AsO$_4$)$_3$[63] are highly sought after for their potential to reveal pristine Weyl fermionic physics. Other typical magnetic Weyl semimetals usually suffer from complex band structures with other trivial pockets or multiple Weyl cones coexisting, such as Co$_3$Sn$_2$S$_2$[64–66] and PrAlGe[67]. Among these candidate ideal Weyl systems, Mn(Bi$_{1-x}$Sb$_x$)$_2$Te$_4$ family stands out due to its highly tunable electronic structure and Fermi level. More importantly, the inherent layered structure with weak van der Waals interactions in Mn(Bi$_{1-x}$Sb$_x$)$_2$Te$_4$ renders it ideally suited for nanofabrication and the creation of thin-film devices through mechanical exfoliation, circumventing the complexities of molecular beam epitaxy synthesis often required for nano devices of other topological materials[58,62]. The combination of intrinsic topological properties, facile synthesis, and device compatibility positions OCVT-grown Mn(Bi$_{1-x}$Sb$_x$)$_2$Te$_4$ as a highly promising platform for advancing topological physics and realizing next-generation electronic and spintronic devices.

In conclusion, through a synergistic approach combining first-principles calculations and experimental investigations, we have definitively demonstrated the critical role of antisite defects in dictating the magnetic and topological properties of FM Mn(Bi$_{1-x}$Sb$_x$)$_2$Te$_4$. Our theoretical calculations reveal that both excessive antisite defects and overmuch Sb doping lead to significant degradation of the Weyl cone in the field-forced FM phase. To overcome this challenge, we successfully developed an OCVT method, enabling the synthesis of stoichiometric Mn(Bi$_{1-x}$Sb$_x$)$_2$Te$_4$ crystals with minimized antisite density. Magnetization measurements corroborate the enhanced crystal quality, evidenced by more intrinsic magnetic ordering and a $p$-type Fermi level shift at a given Sb doping level. The CNP occurs at a minimized Sb-doping level ($x$ ~ 0.20), as confirmed by both ARPES and Hall effect measurements. Our optimized crystals exhibit a record carrier mobility of 2519 cm$^2$/V·s, the highest reported to date. Furthermore, the observation of robust SdH oscillations and pronounced AHE conductivity in the forced FM state, exhibiting strong dependence on carrier type and density, provides compelling evidence for the realization of the type-II Weyl semimetal state. These high-quality Mn(Bi$_{1-x}$Sb$_x$)$_2$Te$_4$ crystals, characterized by low carrier density, enhanced mobility, and robust magnetism, offer an unprecedented experimental platform for exploring emergent quantum phenomena arising from the intricate interplay between band topology and magnetic order. Moreover, the inherent exfoliability of these layered materials unlocks significant potential for future investigations of rich topological phenomena in nanoflake devices, including the quantum anomalous Hall effect, axion insulator state, and Fermi-arc-mediated surface transport, paving the way for novel topological electronic and spintronic applications.

## Methods

**Single crystal growth.** High-quality single crystals were synthesized by the OCVT method. Starting materials, including Mn (piece, 99.98%), Bi (grain, 99.999%), Sb (lump, 99.99%) and Te (lump, 99.99%), were mixed in the molar ratio of MnTe : (Bi, Sb)$_2$Te$_3$ = $y$ : 1, with $y$ selected between 2 and 3. For enhanced reactivity, Mn pieces were ground into fine powders. For Sb-doped crystals, a slightly higher Sb content than the target substitution level $x$ was used, where $x$ = Sb%/(Sb%+Bi%). The raw materials (1 ~ 2 g) were sealed in a quartz ampule under vacuum. The ampule was first heated at 600 ~ 630 °C as a pre-reaction for a uniform mixture. Subsequently, I$_2$ granule of around 20 mg was added with the precursors as the transport agent. The re-sealed ampule was then placed into a two-zone tube furnace (OTF-1200X-II, Kejing) with a slight elevation at the cold end, and exposed to a temperature difference of 20 °C between the hot and cold zones. For higher Sb-doping levels, the temperature difference was increased to 30 °C. The reaction temperature was adjusted depending on the doping level (details in Supplementary Note 4). After a two-week reaction and natural cooling, shiny crystals measuring several millimeters were obtained at the cold end.

**Component and structural characterization.** Elemental composition and atomic percentages of the single crystals were determined by energy dispersive spectroscopy (EDX, Oxford X-Max), utilizing an electron beam energy of 10 keV and an integral spectrum energy range of 20 keV. X-ray diffraction (XRD) measurements were carried out using a diffractometer (Bruker, D8 Discover) equipped with Cu K$_\alpha$ radiation ($\lambda$ = 1.5418 Å). A 2-bounce crystal monochromator was installed in the beam path to remove the K$_{\alpha 2}$ component and get a monochromatic K$_{\alpha 1}$ X-ray ($\lambda$ = 1.54056 Å) irradiating on the (001) surface of the crystals. The scanning transmission electron microscopy (STEM) imaging was performed on a double aberration corrected field emission transmission electron microscope (Themis Z, Thermo Fisher Scientific) operated at 300 kV. The probe forming semi-convergent angle is 21.4 mrad, and the semi-collection angle of High-angle annular dark field (HAADF) detector is 79–200 mrad. Scanning tunneling microscopy (STM) was performed on an OCVT-grown Mn(Bi$_{0.8}$Sb$_{0.2}$)$_2$Te$_4$ single crystal. The (001) surface, prepared by *in situ* cleavage under a pressure of 1.9×10$^{−10}$ mbar, was imaged in atomic resolution with a bias voltage of −0.75 V and tunneling current of 230 pA at 77 K.

**Magnetic measurements**. Temperature- and field-dependent magnetization measurements were conducted using a vibrating sample magnetometer (VSM) equipped in a physical property measurement system (PPMS, EverCool II, Quantum Design, ±9 T) and VSM mode in SQUID (MPMS3, Quantum Design, ±7 T). Standard quartz sample holders were used. For temperature-dependent magnetization, an external field of 500 Oe was applied under both zero-field-cooled (ZFC) and field-cooled (FC) conditions.

**ARPES measurements.** The ARPES measurements were performed at beamline BL13U of the National Synchrotron Radiation Laboratory (NSRL) in Hefei, China. The samples were cleaved and measured *in situ* under ultra-high vacuum conditions with a base pressure better than 6 × 10$^{-11}$

Torr, at a temperature of 10 K. The photon energy was set to $h\nu = 13$ eV, which allowed for clear detection of the band dispersions, and the energy resolution was better than 20 meV.

**Electrical transport measurements.** Electrical transport measurements were performed in a commercial variable temperature insert with a Cryogenic superconducting magnet (1.6 ~ 300 K, ±12 T). Clean, flat surfaces were cleaved from the as-grown bulk crystals, and aluminum wires were bonded to the surface using silver paste (DuPont, 4929N) to form a standard six-wire Hall configuration. A voltage-controlled current source (Stanford Research System, CS580) was used to provide an alternating current at 17.777 Hz, and the voltage signals were recorded by several lock-in amplifiers (Stanford Research System, SR865).

**Theoretical calculations.** The first-principles calculations were performed using the projector augmented-wave (PAW)[68] pseudopotential method as implemented in the VASP code[69]. The Perdew-Burke-Ernzerhof generalized gradient approximation (GGA-PBE)[70] is employed to describe the exchange and correlation functional. The plane-wave energy cutoff was set to 550 eV, with a total energy convergence criterion of $10^{-6}$ eV. All atoms in the unit cell were allowed to relax until the Hellmann-Feynman force on each atom was less than 0.001 eV/Å. The GGA+U method[71] with an effective Hubbard U value of 3.0 eV was used to address the correction effects of $d$ electrons of Mn atoms[6]. The Brillouin zone integral is implemented on a Γ-centered grid mesh of $9 \times 9 \times 5$. Supercells that correspond to $2 \times 2 \times 1$ of the original unit cells of $MnBi_2Te_4$ and $Mn(Bi_{1-x}Sb_x)_2Te_4$ were used to simulate the influence of magnetic antisites or Sb substitutions at various densities.

## Data availability
The Source Data underlying the figures of this study are available with the paper. All raw and derived data supporting the findings in the study are available from the corresponding author upon reasonable request.

## Acknowledgments

C.Z. was sponsored by the National Key R&D Program of China (Grant No. 2022YFA1405700), the National Natural Science Foundation of China (Grant No. 92365104, 12174069), and the Shanghai Pilot Program for Basic Research-Fudan University 21TQ1400100 (25TQ001). L.M. was sponsored by the National Natural Science Foundation of China (Grant No. 12374058). T.Z. was sponsored by the National Natural Science Foundation of China (Grants No. 12474155), and the Zhejiang Provincial Natural Science Foundation of China (LR25A040001). X.Y. was sponsored by the National Key R&D Program of China (Grant No. 2023YFA1407500), Scientific Research Innovation Capability Support Project for Young Faculty (Grant No. ZYGXQNJSKYCXNLZCXM-M11), the National Natural Science Foundation of China (Grant No. U24A2012, 12174104). C.-L.Z. was sponsored by the National Natural Science Foundation of China (Grant No. 62171136).

**Author Contributions Statement**

C.Z. conceived the ideas and supervised the overall research. H.C. developed the OCVT method and synthesized high-quality $Mn(Bi_{1-x}Sb_x)_2Te_4$ crystals. H.C. and J.W. performed the electrical transport measurements with the help of Y.W., Z.X., Y.X., Y.M., and J.G.. X.D. and T.Z. performed the first-principles calculations. H.L. and L.M. performed the ARPES and STM characterizations. H.C. carried out the magnetization measurements with the help of Z.J., X.C., Y.M., and F.X.. W.H. and C.-L.Z. performed the STEM measurements. H.C., J.W. and C.Z. analyzed and interpreted the results with the assistance of X.J., P.L., F.Z., and X.Y.. H.C., J.W. and C.Z. wrote the paper with assistance from all other co-authors.

**Competing Interests Statement**

The authors declare no competing interests.

**Tables**

| $x$ | $M_{sat}$ | $Mn_{EDX}$ | $(Bi, Sb)_{EDX}$ | $Mn_{Mn}$ (%) | $Mn_{Bi/Sb}$ (%) | $Bi/Sb_{Mn}$ (%) |
|---|---|---|---|---|---|---|
| 0 | 3.99 | 0.99 | 2.05 | 92.44 | 3.28 | 7.56 |
| 0.04 | 3.95 | 0.99 | 2.03 | 92.01 | 3.50 | 7.99 |
| 0.12 | 3.86 | 0.98 | 2.03 | 90.12 | 3.94 | 9.88 |
| 0.16 | 3.78 | 0.98 | 2.01 | 89.27 | 4.37 | 10.73 |
| 0.18 | 3.69 | 0.98 | 2.06 | 88.31 | 4.85 | 11.69 |
| 0.20 | 3.67 | 0.98 | 2.03 | 88.09 | 4.95 | 11.91 |
| 0.21 | 3.63 | 1.01 | 2.01 | 90.35 | 5.32 | 9.65 |
| 0.24 | 3.60 | 0.98 | 2.03 | 87.35 | 5.33 | 12.65 |
| 0.26 | 3.51 | 1.00 | 2.06 | 88.15 | 5.92 | 11.85 |

| $x$ | Bulk | Growth | Ref | Methods | $Mn_{Bi/Sb}$ (%) | $Bi/Sb_{Mn}$ (%) |
|---|---|---|---|---|---|---|
| 0 | $MnBi_2Te_4$ | Bridgman method | 21 | STM, XRD | 3.4–5.2 | 11 |
| 0.24 | $Mn(Bi_{1-x}Sb_x)_2Te_4$ | Solid reaction | 25 | Neutron | 6 | 18 |
| 1 | $MnSb_2Te_4$ | Self-flux | 22 | Neutron/EDX | 15.80 | 32.60 |

**Table 1. A comprehensive comparison of saturation magnetization, element ratio and defects concentration for $Mn(Bi_{1-x}Sb_x)_2Te_4$ samples with varying Sb content $x$.** The saturation

magnetizations $M_{sat}$ are extracted from the magnetization values at 9 T and 2 K in Fig. 3b. The element atomic ratios of Mn and (Bi+Sb) come from the normalized results of EDX, denoted as $Mn_{EDX}$ and $(Bi, Sb)_{EDX}$ with $Te_{EDX}$ set to a reference value of 4. $Mn_{Mn}\%$ corresponds to the amount percentage of Mn atoms in the main Mn layer, while $Mn_{Bi/Sb}\%$ and $Bi/Sb_{Mn}\%$ refer to the amount percentage of antisite defects in the single atomic layer. The deviations between $Mn_{Bi/Sb}\%$ and half of $Bi/Sb_{Mn}\%$ primarily stem from compositional non-stoichiometry in $Mn(Bi_{1-x}Sb_x)_2Te_4$ crystals, with Mn:(Bi+Sb) ratio deviating from 1:2. Compared with the defect level of samples in previous works, we can conclude a lower level of antisite defects and higher quality in OCVT samples.

| Ref | Bulk | Maximum mobility ($cm^2/V \cdot s$) |
|---|---|---|
| 14 | $MnBi_2Te_4$ | 74 (at 1.6 K) |
| 5 | $MnBi_2Te_4$ | 100 (at 5 K) |
| This work | $MnBi_2Te_4$ | 282 (at 2 K) |
| Ref | Bulk | Maximum mobility ($cm^2/V \cdot s$) |
| 9 | $Mn(Bi_{1-x}Sb_x)_2Te_4$ | 715 (at 75 K) |
| 8 | $Mn(Bi_{1-x}Sb_x)_2Te_4$ | 1071 (at 75 K) |
| 7 | $Mn(Bi_{1-x}Sb_x)_2Te_4$ | 1250 (at 2 K) |
| This work | $Mn(Bi_{1-x}Sb_x)_2Te_4$ | 2519 (at 2 K) |

**Table 2. Comparison of the maximum carrier mobility of $MnBi_2Te_4$ and $Mn(Bi_{1-x}Sb_x)_2Te_4$ samples.** Both the pristine and Sb-doped $MnBi_2Te_4$ in this work show higher mobility than that of previous studies.

**Figure Legends/Captions**

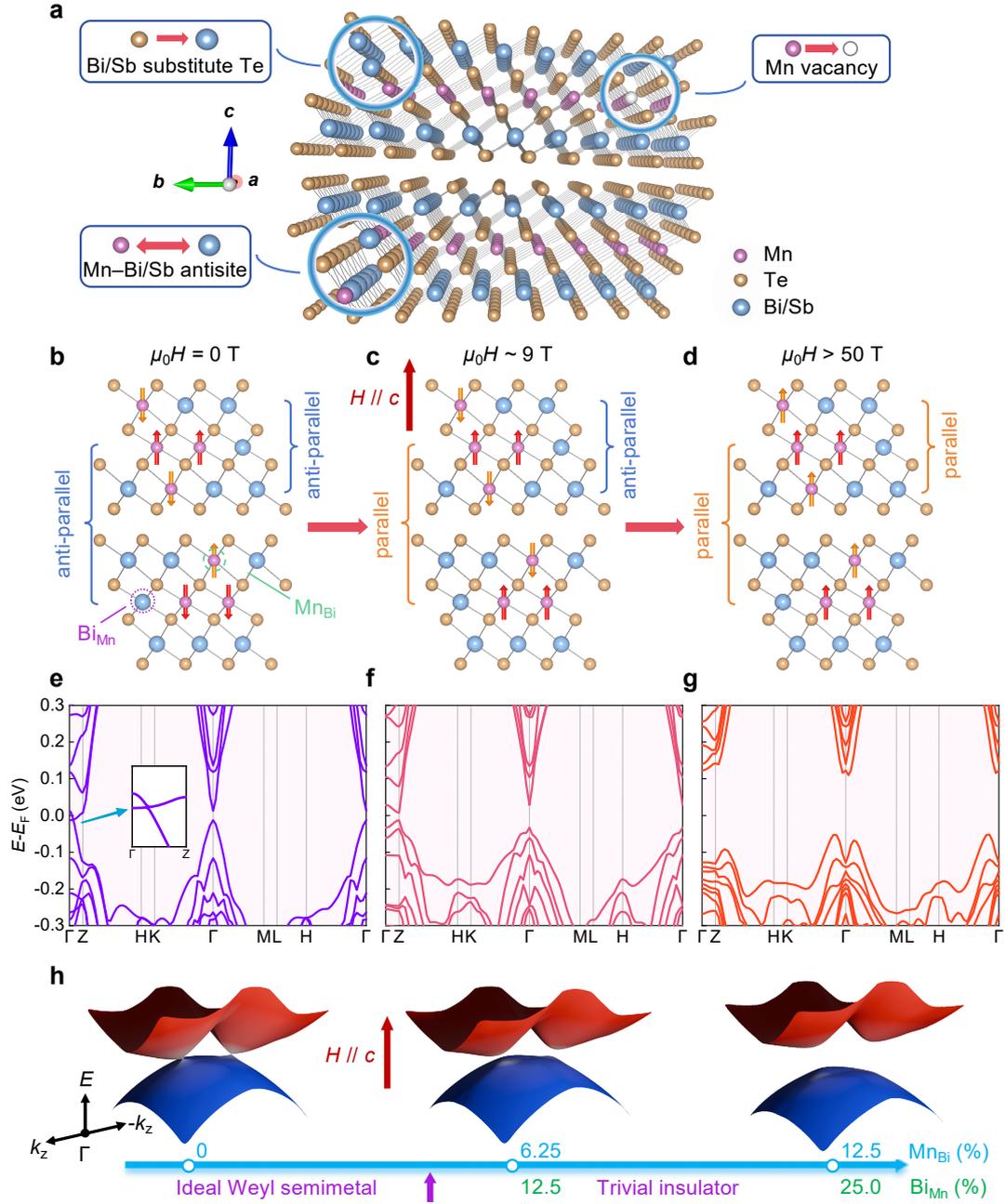

**Fig. 1 Schematics of defect-included Mn(Bi$_{1-x}$Sb$_x$)$_2$Te$_4$ crystal structure, magnetic order and band topology in its ferromagnetic state.**

**a** Schematic of Mn(Bi$_{1-x}$Sb$_x$)$_2$Te$_4$ crystal structure with several typical types of defects, including Mn-Bi/Sb antisites, Bi/Sb atoms occupying Te sites, and vacancies at Mn sites. Two septuple layers are shown for clarity.

**b, c, d** Magnetic structures in Mn(Bi$_{1-x}$Sb$_x$)$_2$Te$_4$ with the coexisted Mn$_{Bi/Sb}$ and Bi/Sb$_{Mn}$ defects under varying magnetic fields. Without magnetic field in (**b**), interlayer Mn$_{Mn}$ moments form an antiferromagnetic order, while within a SL, Mn$_{Bi/Sb}$ antisites exhibit an antiparallel arrangement with Mn$_{Mn}$ moments. With a moderate field of 9 T along the $c$ axis as (**c**), the sublattices of Mn$_{Mn}$ are forced into a ferromagnetic state, while the sublattices of Mn$_{Mn}$ and Mn$_{Bi/Sb}$ remain antiferromagnetically coupled. At higher fields exceeding 50 T in (**d**), Mn moments in all layers align parallel to the applied field, achieving a fully saturated ferromagnetic state.

**e** Band structure of FM MnBi$_2$Te$_4$ with idealized Mn moments alignment, assuming no antisites. Inset is the enlarged view of the Weyl point located along the Γ–Z line.

**f, g** Band structure of MnBi$_2$Te$_4$ under interlayer FM and intralayer AFM orders. The calculation model contains 6.25% Mn$_{Bi}$ and 12.5% Bi$_{Mn}$ defects in (**f**), and 12.5% Mn$_{Bi}$ and 25% Bi$_{Mn}$ defects in (**g**).

**h** A schematic diagram of the defect-induced Weyl cone degeneration. The blue arrow represents the increasing density of Mn-Bi antisites, transitioning the ideal Weyl semimetal into a trivial insulator. The topological phase transition may occur at a certain level of antisite density, illustrated by the purple arrow.

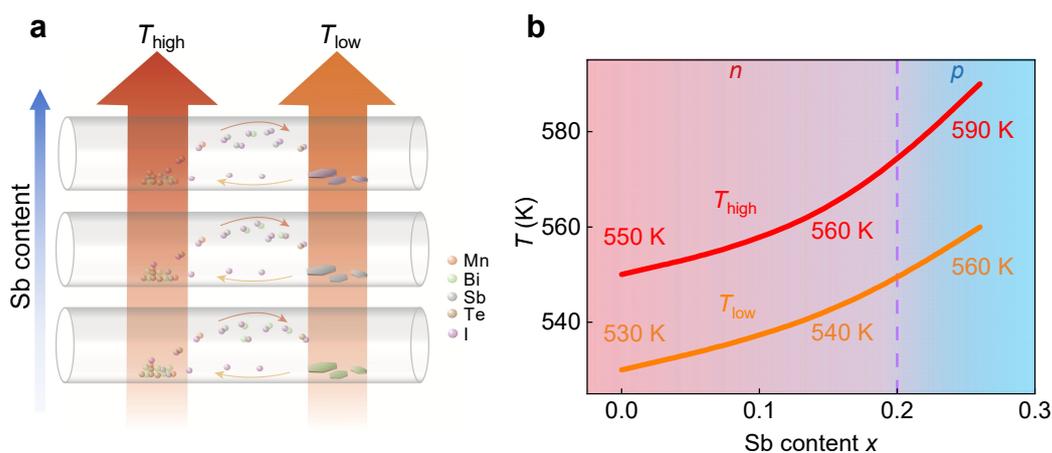

**Fig. 2 Schematic images of the OCVT method.**

**a** An illustration of the OCVT method. As the Sb content increases, higher temperatures are applied at both the hot and cold zones ($T_{high}$ and $T_{low}$) of the quartz ampules. All the raw materials utilized in synthesis are denoted by balls with different colors.

**b** Optimized temperatures at the hot and cold zones for growing high-quality Mn(Bi$_{1-x}$Sb$_x$)$_2$Te$_4$ crystals, tailored to different Sb contents.

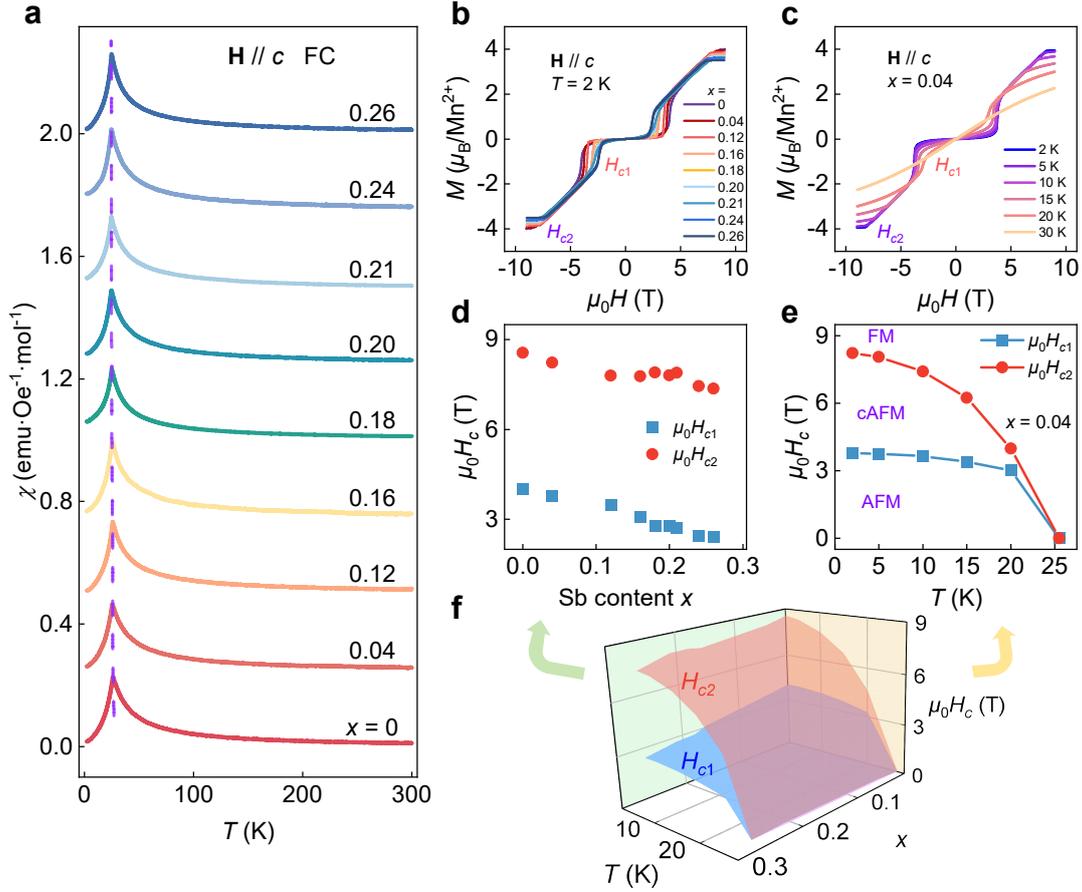

**Fig. 3 Magnetism characterization of Mn(Bi$_{1-x}$Sb$_x$)$_2$Te$_4$.**

**a** Temperature dependence of magnetic susceptibility for various Sb content $x$. The curves are measured in FC condition with a field of 500 Oe applied along the $c$ axis. The values of $T_N$ are marked with a purple dashed line.

**b** Magnetic field dependence of magnetization for Sb content ranging from $x = 0$ to 0.26 at 2 K.

**c** Magnetic field dependence of magnetization for Sb content $x = 0.04$ at different temperatures.

**d, e, f** Critical fields for the spin-flop transition and saturation ($H_{c1}$ and $H_{c2}$) extracted from $M(H)$ curves by identifying the d$M$/d$H$ peaks and d$^2M$/d$H^2$ valleys, respectively. The variations of $H_{c1}$ and $H_{c2}$ with temperature and Sb content $x$ are illustrated in a three-dimensional diagram (**f**). The X-Z cut (**d**) and Y-Z cut (**e**) of the 3D figure show the dependence of $H_{c1}$ and $H_{c2}$ on Sb content $x$ and the magnetic transitions at different temperatures.

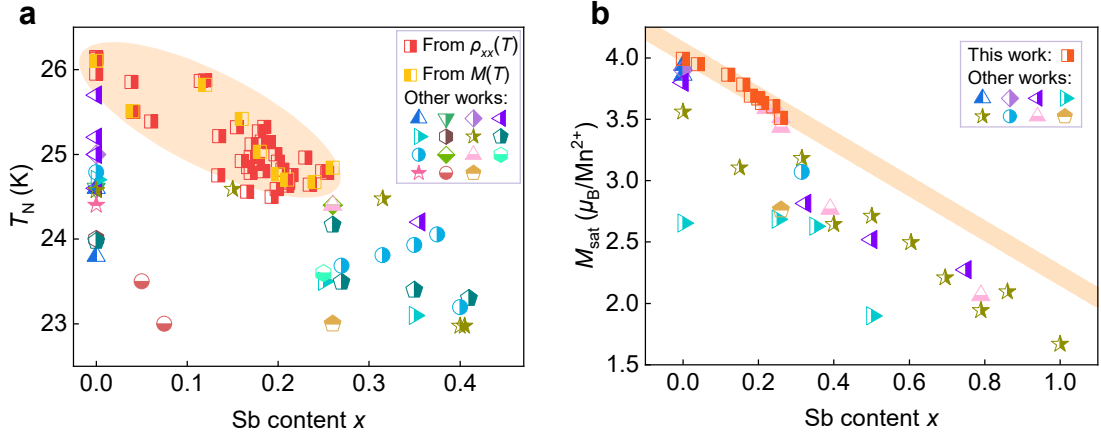

**Fig. 4 Comparison of magnetic properties with previous studies.**
**a** Comparison of Néel temperatures $T_N$ across various studies, with data collected from this work emphasized within the orange ellipse. The yellow and red squares represent $T_N$ extracted from $M(T)$ and $\rho_{xx}(T)$ curves, respectively. The corresponding references to the data extracted from other works are refs. 39, 36, 26, 42, 29, 24, 34, 52, 10, 9, 8, 43, 27, 49, 53, sequencing from top left to bottom right in figure key.
**b** Comparison of saturation magnetization $M_{sat}$ as a function of Sb content $x$. The $M_{sat}$ values are extracted from the saturation magnetization from the $M(H)$ curves at a moderate field (~ 9 T). The orange line represents a linear fitting of saturation magnetization in this work. The significantly higher saturation magnetization and Néel temperature observed in our Mn(Bi$_{1-x}$Sb$_x$)$_2$Te$_4$ crystals indicate a more intrinsic magnetism with high quality and fewer magnetic defects. The corresponding references to the data extracted from other works are refs. 39, 26, 42, 29, 34, 10, 8, 53, sequencing from top left to bottom right in figure key.

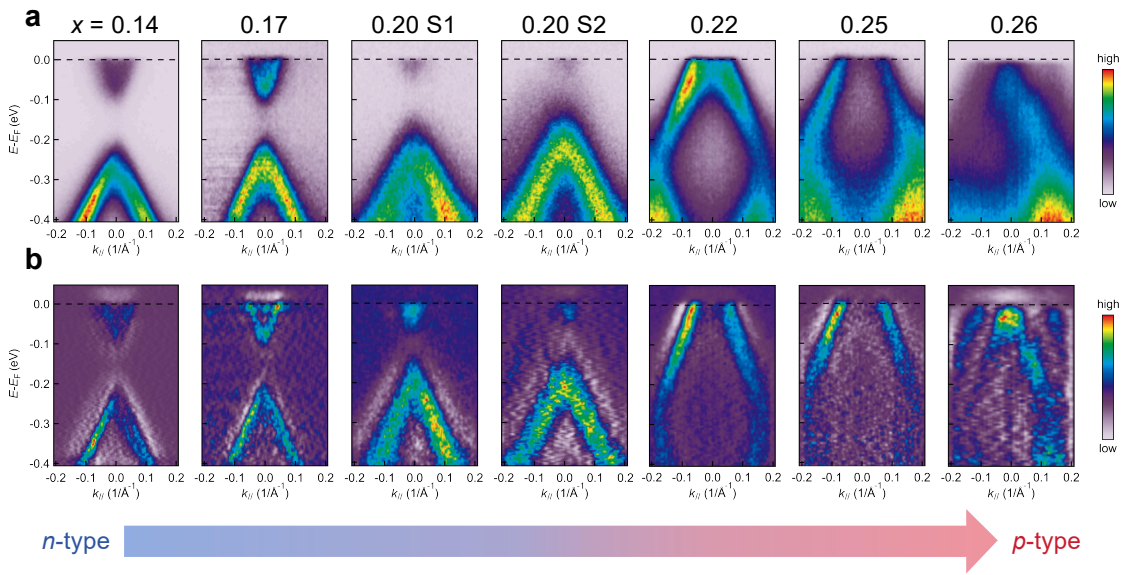

**Fig. 5 ARPES results for Mn(Bi$_{1-x}$Sb$_x$)$_2$Te$_4$ samples.**
**a, b** Raw and second derivative spectra of ARPES measurements for Mn(Bi$_{1-x}$Sb$_x$)$_2$Te$_4$ samples with $x$ values of 0.14, 0.17, 0.20, 0.22, 0.25, and 0.26. Two samples with $x = 0.20$ were measured to confirm the band-edge conduction feature. The Fermi level, indicated by the black dashed line, shifts downwards from the conduction band to the valence band, residing within the gap at $x \sim 0.20$, suggesting the carrier transition from *n*-type to *p*-type as Sb doping content *x* increases.

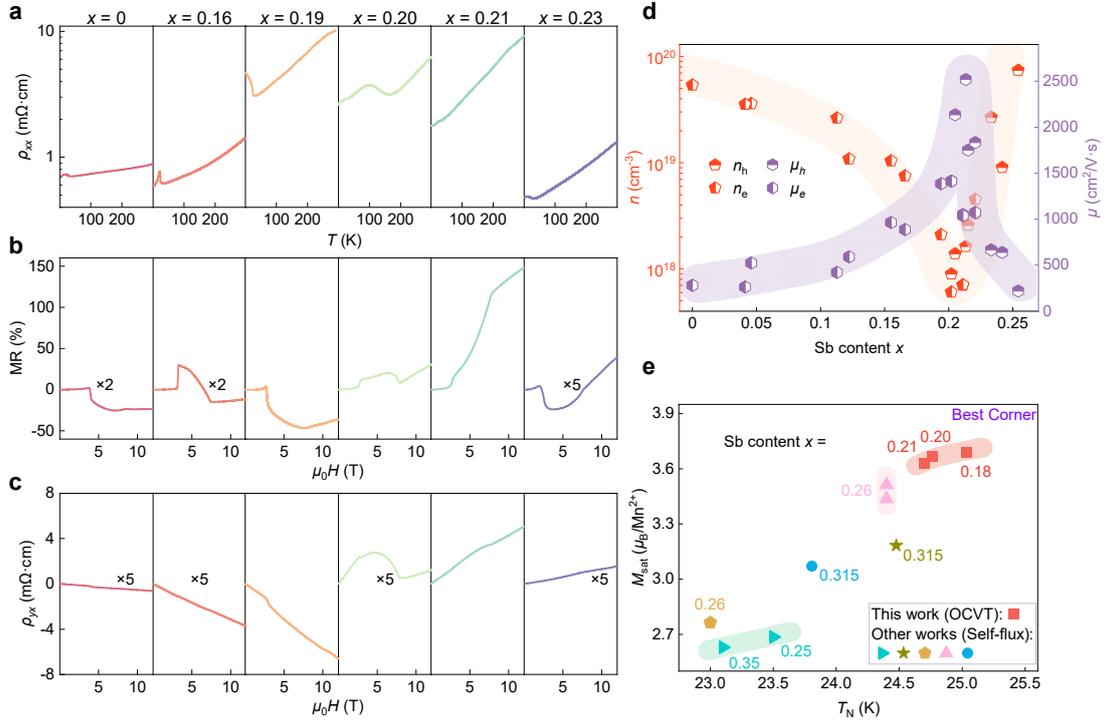

**Fig. 6 Electrical transport properties of Mn(Bi$_{1-x}$Sb$_x$)$_2$Te$_4$.**

**a** Temperature-dependent longitudinal resistivity $\rho_{xx}(T)$ with varying Sb content *x*. All curves are measured with the current flowing in the *ab* plane.

**b** Field dependence of magnetoresistance MR for different Sb content at 2 K.

**c** Field dependence of Hall resistivity $\rho_{yx}(H)$ for different Sb content at 2 K. In field-dependent measurements, the field is applied along the *c* axis while the current flows in the *ab* plane. For clarity, part of the curves in panels (**b**) and (**c**) are scaled by factors, as marked in the figure.

**d** Evolution of carrier density and mobility as a function of *x* for selected samples with high mobility. The carrier mobility approaches its maximum as carrier density nears a minimum, indicating enhanced electrical performance around the charge neutral point. The carrier density of electrons and holes, *n* and *p*, are plotted in log scale.

**e** A comparative comparison of saturation magnetization $M_{sat}$ and Néel temperature $T_N$ for samples near the CNP in both our work and previous studies. The referenced magnetic properties from the literature correspond to refs. 29, 34, 53, 8, 10, respectively, as indicated from left to right in the figure key. In this work, samples from OCVT exhibit a charge neutral point at $x \sim 0.20$. Notably, since the magnetization data for samples with CNP compositions are not included in ref. 29 ($x_{CNP}$ = 0.30) and ref. 10 ($x_{CNP}$ = 0.35), we selected magnetic properties of samples with the closest doping content near the CNP from these references to enable meaningful comparison.

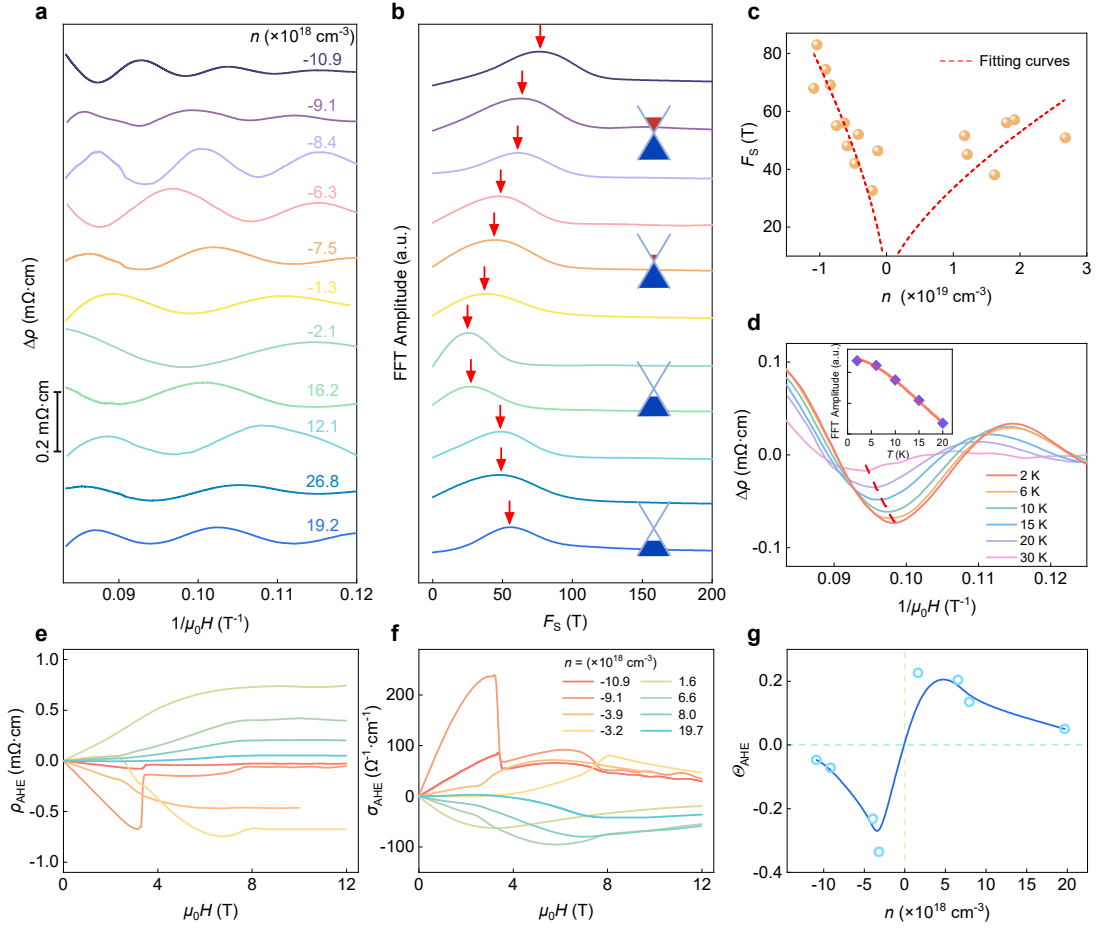

**Fig. 7 Shubnikov-de Haas (SdH) oscillations and anomalous Hall effect (AHE) in Mn(Bi$_{1-x}$Sb$_x$)$_2$Te$_4$.**

**a** SdH oscillations for samples with varying carrier densities (marked beside the corresponding curves, from *n*-type to *p*-type) at 2 K with magnetic field along the *c* axis. The scale factors for each curve are ×2, ×20, ×10, ×5, ×20, ×5, ×1, ×20, ×50, ×50, and ×20, respectively, from top to bottom.

**b** Fast Fourier transform (FFT) spectra of the SdH oscillations in (**a**). The oscillation frequencies (red arrows) are almost consistent with the Fermi surface evolution from *n*-type to *p*-type, as illustrated on the side. Curves of the same color in (**b**) and (**a**) correspond to identical carrier densities.

**c** SdH oscillation frequencies as a function of carrier density. The red dashed lines represent Fermi surface shape fitting curves for the electron and hole regions.

**d** Temperature dependence of SdH oscillations amplitude from 2 K to 30 K of a sample with an electron density of -2.1×10$^{18}$ cm$^{-3}$. The red dashed line indicates the shift of the peak positions with increasing temperature. The inset shows the fitting of the effective mass using the Lifshitz-Kosevich (LK) formula, based on the temperature-dependent FFT amplitude.

**e** Field-dependent anomalous Hall resistivity $\rho_{AHE}$ in samples with various carrier densities, where $\rho_{AHE} = \rho_{yx} - R_H \mu_0 H$ ($R_H$ is the linear slope of $\rho_{yx}$ versus magnetic field in the FM state).

**f** Field-dependent anomalous Hall conductivity $\sigma_{AHE}$ among the same batch of samples in (**e**). $\sigma_{AHE}$ is defined as $\sigma_{AHE} = \rho_{AHE}/(\rho_{xx}^2 + \rho_{yx}^2)$.

**g** Anomalous Hall angle $\Theta_{AHE} = \sigma_{AHE}/\sigma_{xx}$ as a function of carrier density. All data points are obtained at 2 K and 10 T.

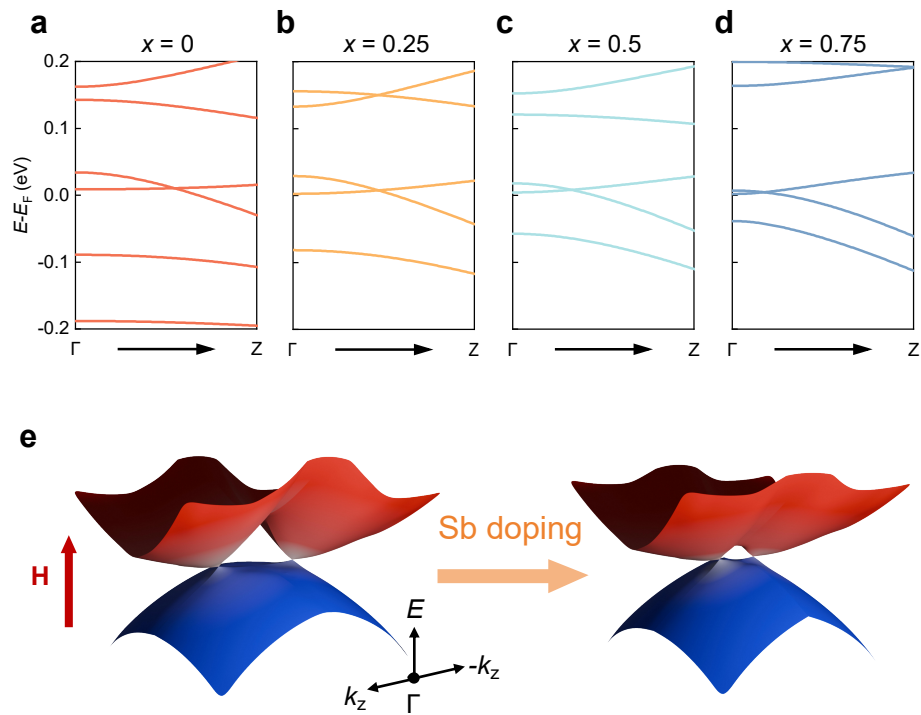

**Fig. 8 Theoretical calculations of band structure degradation from Sb-doping.**
**a-d** The band structure along the Γ-Z line for FM Mn(Bi$_{1-x}$Sb$_x$)$_2$Te$_4$ with Sb content $x$ = 0, 0.25, 0.5, 0.75 for (**a**), (**b**), (**c**) and (**d**), respectively. As Sb content increases, the Weyl point shifts towards the Γ point with a reduced Weyl point separation Δ$k$ in momentum space.
**e** A schematic diagram of the Weyl cone evolution caused by Sb-doping, with the Weyl points shrinking.